\documentclass[twocolumn,prl]{revtex4}

\usepackage{graphicx}

\begin{document}

\title{Infinite Randomness Fixed Points for
Chains of Non-Abelian Quasiparticles}

\author{N.\ E.\ Bonesteel and Kun Yang }

\affiliation{Department of Physics and National High Magnetic
Field Laboratory, Florida State University, Tallahassee, Florida
32310, USA}

\pacs{}

\begin{abstract}
One dimensional chains of non-Abelian quasiparticles described by
$SU(2)_k$ Chern-Simons-Witten theory can enter random singlet
phases analogous to that of a random chain of ordinary spin-1/2
particles (corresponding to $k \rightarrow \infty$).  For $k=2$
this phase provides a random singlet description of the infinite
randomness fixed point of the critical transverse field Ising
model. The entanglement entropy of a region of size $L$ in these
phases scales as $S_L \simeq \frac{\ln d}{3} \log_2 L$ for large
$L$, where $d$ is the quantum dimension of the particles.
\end{abstract}

\maketitle

A particularly exotic form of quantum order is possible in two
space dimensions --- so-called \emph{non-Abelian} order
\cite{mooreread}. In states with non-Abelian order, when certain
localized quasiparticle excitations are present there is a
low-energy Hilbert space whose dimensionality grows exponentially
with the number of these quasiparticles.  When these
quasiparticles are well separated, this low-energy space becomes
degenerate, and its states are characterized by purely topological
quantum numbers, meaning they cannot be distinguished by local
measurements. If these quasiparticles are then adiabatically moved
around one another, unitary transformations corresponding to
non-Abelian representations of the braid group are carried out on
this degenerate space. Aside from their intrinsic scientific
interest, recent attention has focused on the possibility of one
day using non-Abelian states to perform fault-tolerant quantum
computation \cite{kitaev,freedman}.

Recently Feiguin et al.~\cite{feiguin} have studied models of {\it
interacting} non-Abelian quasiparticles, specifically uniform
chains in which neighboring quasiparticles are close enough
together to lift the degeneracy of the topological Hilbert space.
In this Letter we study a related class of {\it random}
interacting chains of non-Abelian quasiparticles. We are motivated
both by \cite{feiguin} and by recent work of Refael and Moore
\cite{refael, refael2} showing that the entanglement entropy of
certain random one-dimensional models scales logarithmically with
a universal coefficient.  We find the same is true here for an
infinite class of models.

Exact diagonalization studies \cite{morf,rezayihaldane,yang}
provide compelling evidence that the experimentally observed
$\nu=5/2$ fractional quantum Hall (FQH) state is a non-Abelian
state described by the Moore-Read ``Pfaffian" state
\cite{mooreread}. This state belongs to a wider class of
non-Abelian FQH states introduced by Read and Rezayi
\cite{readrezayi}, labeled by index $k$. In this class, the $k=1$
state is an ordinary (Abelian) Laughlin state, the $k=2$ state is
the Moore-Read state, and all subsequent integer $k$ values
describe new non-Abelian states. There is some numerical evidence
\cite{readrezayi,rezayiread} that the $k=3$ Read-Rezayi state
describes the experimentally observed $\nu = 12/5$ FQH state
\cite{xia}.

\begin{figure}[t]
\centerline{\includegraphics[scale=.33]{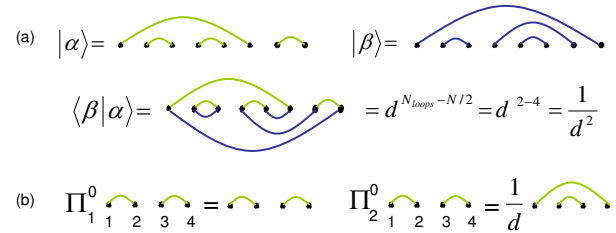}~~~~~~~~~~~~}
\caption{(a) Two non-crossing singlet states for $SU(2)_k$
particles and their overlap. (b) Action of the singlet projection
operators $\Pi_1^0$ (which acts on particles 1 and 2) and
$\Pi_2^0$ (which acts on particles 2 and 3) on a particular
non-crossing singlet state. The quantity $d$ appearing in (a) and
(b) is the quantum dimension of the particles.} \label{vb}
\end{figure}

The quasiparticle excitations of the Read-Rezayi states with index
$k$ can be viewed (up to Abelian phases) as particle excitations
in $SU(2)_k$ Chern-Simons-Witten theory \cite{slingerlandbais}.
These particles are characterized by their topological charge, a
quantum number which can be viewed as a ``$q$-deformed" spin
\cite{fuchs}. At level $k$, topological charge can take the values
$0,\frac{1}{2},1,\cdots,\frac{k}{2}$, and obeys the fusion rule,
\begin{eqnarray}
s_1 \otimes s_2 = |s_1 - s_2| \oplus  \cdots \oplus
 \min(s_1 + s_2, k - s_1 - s_2).\label{fusion}
\end{eqnarray}
For $k\ge 2$ this implies $\frac{1}{2} \otimes \frac{1}{2} = 0
\oplus 1$. Thus, when combining two particles with topological
charge 1/2 the resulting state can either have topological charge
0 or 1. For ordinary spin-1/2 particles the former would be
referred to as a singlet and the latter as a triplet.  We will use
the same terminology for $SU(2)_k$ particles, though it should be
noted that here there is no $S_z$ degeneracy, i.e. there is only
one ``triplet" state. (For reviews of the general theory of
non-Abelian particles see \cite{preskill,kitaev2}).

The total spin 0 sector of a one-dimensional chain of ordinary
spin-1/2 particles is spanned by the the set of all
``non-crossing" singlet states, i.e. states in which pairs of
particles form singlet bonds in such a way that these bonds do not
cross (see Fig.~\ref{vb}(a)). Furthermore, these non-crossing
states are linearly independent \cite{rumer}, and their number,
and hence the dimensionality of the spin 0 Hilbert space, grows
asymptotically as $2^N$ for large $N$.

Using the generalized notion of singlet described above,
non-crossing singlet states can also be used as a basis for the
total topological charge 0 sector of a one-dimensional chain of
$SU(2)_k$ particles \cite{notation}. In this case the
interpretation is that any pair of particles connected by a
singlet bond will fuse to topological charge 0 if brought together
\cite{noncrossing}.

For $N$ ordinary spin-1/2 particles, the overlap of two
non-crossing singlet states $|\alpha\rangle$ and $|\beta\rangle$
can be computed by overlaying the two bond configuration and
counting the number of closed loops, $N_{loops}$. The overlap is
then given by $\langle \alpha | \beta \rangle = 2^{N_{loops} -
N/2}$. For $SU(2)_k$ particles this overlap rule becomes $\langle
\alpha | \beta \rangle = d^{N_{loops} - N/2}$ where $d = 2 \cos
\frac{\pi}{k+2}$ is the ``quantum dimension" of the particles (see
Fig.~\ref{vb}(a)) \cite{preskill,kitaev2}. For these values of $d$
the non-crossing states are no longer linearly independent
--- they satisfy linear relations known as the
Jones-Wenzl projectors \cite{kaufmann} which reduce the size of
the Hilbert space so that its dimensionality grows asymptotically
not as $2^N$, but as $d^N$ for large $N$.

Consider a random one-dimensional chain of $SU(2)_k$ particles.
Following \cite{feiguin}, we assume that neighboring particles are
close enough together so that the singlet and triplet fusion
channels are split in energy, with the singlet lying lowest. The
Hamiltonian describing this chain is then
\begin{eqnarray}
H = - \sum_i J_i\ \Pi_{i}^{0}, \label{randombond}
\end{eqnarray}
where $J_i > 0$ is the energy splitting associated with particles
at sites $i$ and $i+1$, and $\Pi_{i}^{0}$ is the singlet projection
operator on these particles, the action of which on representative
non-crossing singlet states is shown in Fig.~\ref{vb}(b). The uniform
versions of these models ($J_i = J$) were studied numerically for $k=3$
and analytically for all $k$ in \cite{feiguin}, where they were shown
to be conformally invariant with central charge $c = 1-6/((k+1)(k+2))$.

Because the Hilbert space of this $SU(2)_k$ chain can be described
using a non-crossing singlet basis,  the usual real-space
renormalization group (RG) approach based on decimating singlet
bonds \cite{dasgupta,fisher1} can be straightforwardly applied to
(\ref{randombond}) when the $J_i$'s are random. Each iteration of
this procedure begins by finding the strongest bond in the chain,
i.e. the $J_i$ with the highest value, and making the
approximation that the two particles connected by it fuse to
topological charge 0 and so form a singlet bond.

\begin{figure}[t]
\includegraphics[scale=.35]{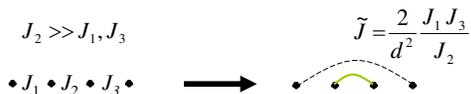}
\caption{One step in the decimation procedure.} \label{decimate}
\end{figure}

The effective interaction $\tilde J$ between the two particles on
either side of this singlet is then determined perturbatively as
follows. Consider four neighboring particles and the associated
three bond strengths $J_1$, $J_2$ and $J_3$, with $J_2 \gg J_1,
J_3$  so that, as described above, a singlet forms between the two
particles connected by $J_2$ (see Fig.~\ref{decimate}). A
straightforward generalization of the usual second-order
perturbation theory calculation for ordinary spin-1/2 particles,
but using the modified overlap rules shown in Fig.~\ref{vb}, then
yields,
\begin{eqnarray}
\tilde J = (2/d^2) J_1 J_3/J_2. \label{pert}
\end{eqnarray}
Provided $d \ge \sqrt{2}$, which is the case for all $k \ge 2$
considered here, $\tilde J$ will always be less than the strength
of the decimated bond $J_2$.  Thus, as this procedure is iterated,
high-energy bonds are systematically eliminated, leading
eventually to a single non-crossing singlet state.

The RG flow produced by this decimation procedure can then be
analyzed in the standard way \cite{dasgupta,fisher1}. Introducing
the logarithmic bond strength variables $\beta_i = \ln
(\Omega/J_i)$, where $\Omega$ is the largest remaining bond
strength at any given stage of decimation, (\ref{pert}) can be
written $\tilde\beta = \beta_1 + \beta_3 + \ln (2/d^2)$. Defining
the flow parameter $\Gamma = \ln( \Omega_0/\Omega)$ where
$\Omega_0$ is the largest bond strength at the outset of the
decimation procedure, and ignoring the $\ln(2/d^2)$ (i.e. taking
$\tilde\beta \simeq \beta_1 + \beta_3)$, an approximation which
can be justified a posteriori due to the broad distribution of
$\beta$'s at the fixed-point, an integro-differential equation can
be written down for the bond strength distribution
$P_\Gamma(\beta)$ \cite{dasgupta,fisher1}. This distribution is
defined so that when the flow parameter is $\Gamma$ the fraction
of bonds with logarithmic strength between $\beta$ and
$\beta+d\beta$ is $P_\Gamma(\beta) d\beta$.  As shown by Fisher
\cite{fisher1}, for almost any initial random bond configuration,
the bond strength distribution flows to the infinite randomness
fixed point distribution, $P_\Gamma(\beta) =
e^{-\beta/\Gamma}/\Gamma$. The resulting phase is known as a
random singlet phase.

It follows that the random $SU(2)_k$ chains (\ref{randombond})
flow to random singlet phases for all $k\ge 2$. In the limit
$k\rightarrow\infty$ this phase corresponds to the usual random
singlet phase for ordinary spin-1/2 particles \cite{fisher1}. For
$k=2$ we now show that the resulting phase can be mapped onto the
infinite randomness fixed point of the critical transverse field
Ising model \cite{fisher2}, thus providing a ``random singlet"
description of this fixed point.

We use the fact that $SU(2)_2$ particles can be represented using
Majorana fermions operators $\gamma_i$ \cite{kitaev2} ---
operators which are self conjugate ($\gamma_i^\dagger = \gamma_i$)
and which satisfy the Clifford algebra $\{\gamma_i,\gamma_j\} = 2
\delta_{ij}$. Two Majorana fermions can be combined to form a
usual fermion, so that, e.g., associated with neighboring sites
$i$ and $i+1$ there is a fermion operator $c_{i,i+1}^\dagger =
(\gamma_i + i \gamma_{i+1})/\sqrt{2}$ which satisfies the usual
anticommutation relation $\{c_{i,i+1},c^\dagger_{i,i+1}\} = 1$ and
which anticommutes with any similar fermion operator constructed
out of a different pair of Majorana fermions.  The Fermi mode
associated with this pair can then be occupied (corresponding to
topological charge 1) or unoccupied (corresponding to topological
charge 0). The singlet projection operator is then $\Pi^{0}_{i} =
1 - c_{i,i+1}^\dagger c_{i,i+1}$, which in the Majorana
representation is $\Pi^{0}_i = i \gamma_i \gamma_{i+1}$.

To map the $SU(2)_2$ chain onto the transverse field Ising model
we first group together neighboring pairs of Majorana fermions.
Letting the index $j$ label these pairs, each of which consists of
a right Majorana fermion ($\gamma^R_j$) and a left Majorana
fermion ($\gamma^L_j$), the Hamiltonian (\ref{randombond}) can be
written
\begin{eqnarray}
H = -\sum_{j} h_j\ i \gamma^L_j \gamma^R_j - \sum_j J_j\ i
\gamma^R_j \gamma^L_{j+1}.\label{majorana}
\end{eqnarray}
Here $h_j$ corresponds to the coupling within the $j^{th}$ pair,
and $J_j$ corresponds to the coupling between the rightmost
particle in the $j^{th}$ pair and the leftmost particle in the
$(j+1)^{st}$ pair. The usual Jordan-Wigner transformation (see,
for example, \cite{vidal}), $\gamma^L_j = \sigma_j^{x}
\prod_{k=1}^{j-1} \sigma_k^z$ and $\gamma^R_j = \sigma_j^{y}
\prod_{k=1}^{j-1} \sigma_k^z$, then maps (\ref{majorana}) onto the
random transverse field Ising model,
\begin{eqnarray}
H = \sum_j h_j \sigma_j^{z}  + \sum_j J_j \sigma_j^{x}
\sigma_{j+1}^{x}.
\end{eqnarray}
Because $h_j$ and $J_j$ are drawn from the same distribution, the
model is at its critical point.

\begin{figure}
\centerline{\includegraphics[scale=.30]{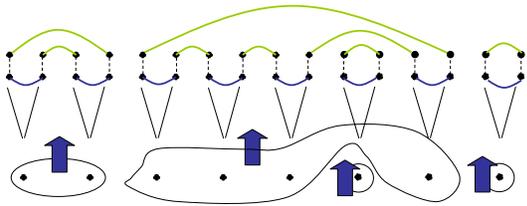}~~~~~~}
\caption{``Random singlet" view of a decimated random transverse
field Ising model.  A random singlet state (green) is overlaid
with a dimer state (blue).  In the dimer state bonds connect pairs
of $SU(2)_2$ particles which are mapped onto the spin-1/2 degrees
of freedom of the transverse field Ising model (bottom row of
dots). Closed loops then correspond to decimated superspins,
indicated by solid arrows.} \label{tfim}
\end{figure}

The usual decimation procedure for the transverse field Ising
model involves two separate steps --- either forming ever larger
``superspins" when the strongest interaction is an Ising
interaction ($J$), or decimating these superspins when the
strongest interaction is a magnetic field strength ($h$)
\cite{fisher2}.  The $SU(2)_2$ ``random singlet" view of this
decimation provides a unified description of these two steps.
Figure \ref{tfim} shows a random singlet state produced by
decimation and a reference ``dimer" state in which bonds connect
pairs of Majorana fermions which correspond to single spins in the
transverse field Ising model. Overlaying these two states produces
closed loops which, in the transverse field Ising model,
correspond to decimated superspins. Essentially, as the decimation
which produces the random singlet state shown in the figure
proceeds, any time a bond forms which does not close a loop this
corresponds to eliminating an Ising interaction and increasing the
number of spins contributing to a superspin. Then, when a bond
forms which closes a loop, the corresponding superspin is frozen
along the direction of the applied field and decimated.

Recently Refael and Moore \cite{refael} have shown that the
entanglement entropy associated with the infinite randomness fixed
points of both the random spin-1/2 Heisenberg chain
($k\rightarrow\infty$) and the transverse field Ising model
($k=2$) have universal scaling properties which can be used to
generalize the notion of central charge to one-dimensional quantum
critical systems which are not conformally invariant. We now show
that the same is true for all the $SU(2)_k$ infinite randomness
fixed points. The entanglement entropy of these states is
calculated by treating a contiguous segment of the chain
consisting of $L$ particles as a subsystem (denoted $A$) of the
full chain. Tracing out the degrees of freedom of the rest of the
chain then yields a reduced density matrix $\rho_A$. The
entanglement entropy is the average over realizations of disorder
of the Von Neumann entropy of this reduced density matrix, $S_L =
-\left\langle {\rm Tr}\ \rho_A \log_2 \rho_A \right\rangle.$

\begin{figure}[t]
\centerline{\includegraphics[scale=.28]{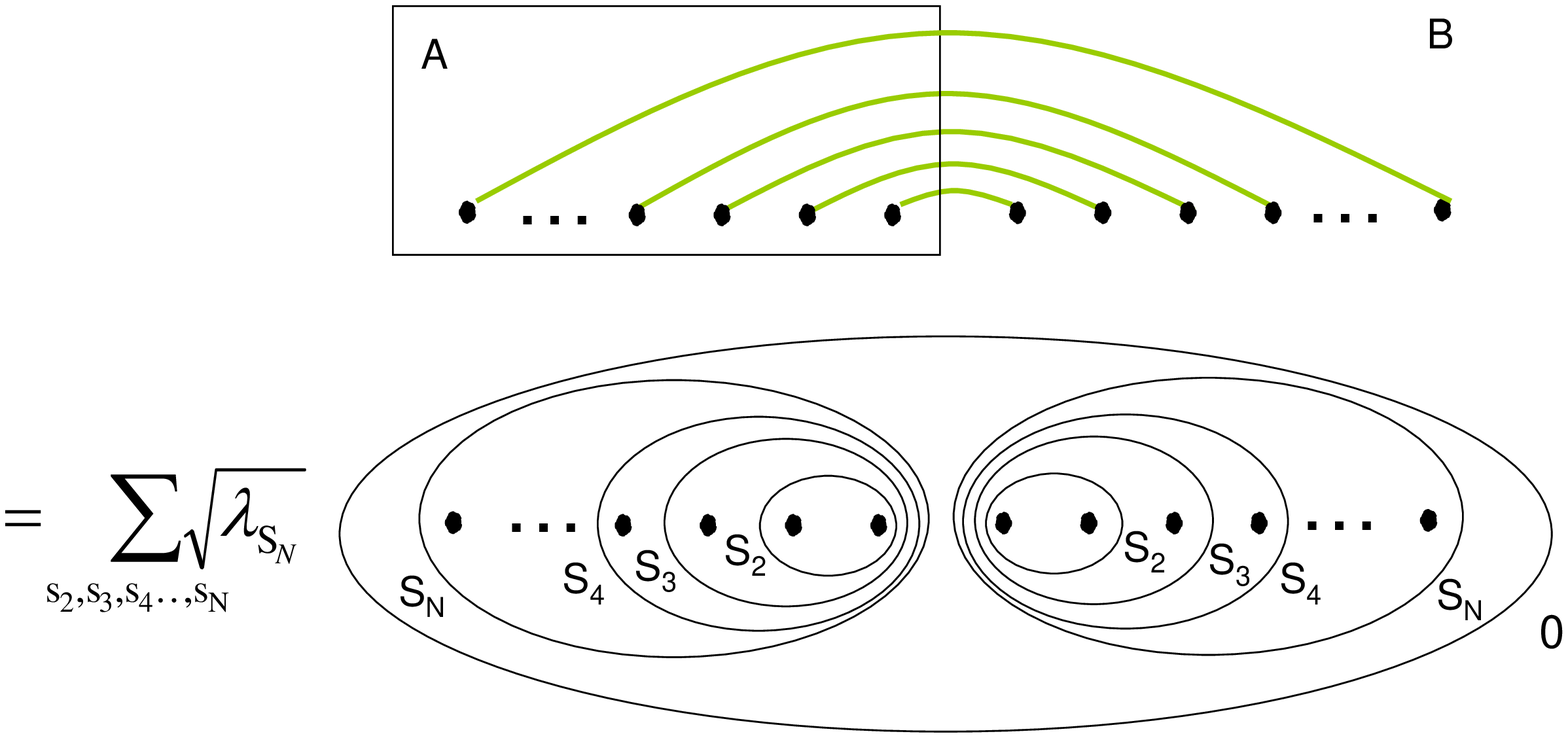}~~~~~~~~~~~~~}
\caption{Schmidt decomposition of a state of $N$ pairs of
$SU(2)_k$ particles connected by singlet bonds. The states in the
decomposition are expressed using a basis in which circles enclose
particles in topological charge eigenstates. The sum is over all
$s_2,s_3\cdots,s_N$ consistent with the fusion rule
(\ref{fusion}).} \label{schmidt}
\end{figure}

In random singlet states the calculation of $S_L$ for large $L$
can be done, as in \cite{refael}, by counting the number of
singlet bonds which connect sites in region $A$ with sites outside
of it, averaging over realizations of disorder, and then
multiplying the result by the entanglement entropy associated with
each bond. All the $SU(2)_k$ random singlet states discussed here
are governed by the same fixed point bond distribution as that
considered in \cite{refael}, so the result of that work that the
average number of bonds contributing to the entanglement scales as
$\frac{1}{3} \ln L$ for large $L$ holds here as well.

To compute the entanglement entropy per bond for $SU(2)_k$
particles, imagine forming $N$ singlet pairs, with one particle
from each pair taken to be in subsystem $A$, the other in
subsystem $B$, as shown in Fig.~\ref{schmidt}.  This figure also
shows a Schmidt decomposition of this state using a basis in which
ovals are drawn around particles in topological charge
eigenstates. The Schmidt coefficients ($\lambda_{s_N}$ in
Fig.~\ref{schmidt}) can be obtained using standard calculation
techniques for non-Abelian particles \cite{preskill,kitaev2}. They
depend only on the total topological charge $s_N$ of the particles
in region $A$ (or equivalently region $B$) of the corresponding
state in the Schmidt decomposition, and are given by
$\lambda_{s_N} = [2s_N+1]/d^N$, where we have introduced the
$q$-integers $[m] = (q^{m/2}-q^{-m/2})/(q^{1/2} - q^{-1/2})$ with
$q = \exp i 2\pi/(k+2)$.

The Von Neumann entropy of the reduced density matrix $\rho_A$
obtained by tracing out the degrees of freedom in region $B$ is
then \cite{nielsen}, $S_A = -\sum_{s_N} {\rm D}(N,s_N)
\lambda_{s_N} \log_2 \lambda_{s_N}$, where ${\rm D}(N,s_N)$ is the
dimensionality of the space of $N$ $SU(2)_k$ particles with total
topological charge $s_N$. Using the fact that, for large $N$,
${\rm D}(N,s_n) \simeq [2s_N + 1] d^N/{\cal D}^2$ where ${\cal
D}^2 = \sum_{s=0}^{k/2} [2s+1]^2$ \cite{preskill,kitaev2}, it
follows that $S_A \simeq N \log_2 d - O(\log_2 k)$ for $N \gg k$.
Thus for large $N$ the entanglement per bond is $\log_2 d$,
reflecting the fact that the size of the Hilbert space of $N$
particles grows asymptotically as $d^N$ \cite{subtle}.

Returning to the $SU(2)_k$ random singlet phases,  multiplying the
average number of bonds leaving a region of size $L$ ($\simeq
\frac{1}{3} \ln L$) by the entanglement per bond ($\simeq \log_2
d$) yields
\begin{eqnarray}
S_L \simeq (\ln d/3) \log_2 L. \label{entropy}
\end{eqnarray}
Following \cite{refael}, if we compare (\ref{entropy}) with the
entanglement entropy of conformally invariant one-dimensional
systems, $S_L \simeq \frac{c}{3} \log_2 L$ where $c$ is the
central charge \cite{holzhey,vidal,calabrese,casini}, it is
natural to define an ``effective central charge" of $\tilde c =
\ln d$ for these phases. In the $k\rightarrow \infty$ limit,
corresponding to the ordinary $SU(2)$ random singlet phase with
$d=2$, we have $\tilde c = \ln 2$, and for $k=2$, corresponding,
as shown above, to the critical point of the random transverse
field Ising model with $d = \sqrt{2}$, we have $\tilde c =
\frac{1}{2} \ln 2$, both of which agree with results obtained in
\cite{refael}.

Finally we note that for the $SU(2)_k$ chains considered here the
effective central charge of the disordered model, $\tilde c = \ln
d$, is always less than the central charge of the uniform model,
$c = 1-6/((k+1)(k+2))$ \cite{feiguin}, though the simple relation
$\tilde c = \ln 2 \times c$ emphasized in \cite{refael} only holds
for $k\rightarrow \infty$ and $k=2$.  This is consistent with the
generalized ``$c$-theorem" envisioned in \cite{refael} which
supposes that the effective central charge decreases along RG
flows between quantum critical points. However, it should be
emphasized that this ``theorem" is not a rigorous result. In
particular, Santachiara \cite{santachiara} has shown that it is
violated by RG flows from the uniform to disordered phases of the
$Z_n$ parafermionic Potts model for $n \ge 42$.

\acknowledgements We thank the KITP at UCSB for its hospitality
during the initial stages of this work, and acknowledge E.
Ardonne, A. Feiguin, A.W.W. Ludwig, J.E. Moore, G. Refael, S.H.
Simon,  J. Slingerland, M. Troyer, and Z. Wang for useful
discussions. This work was supported by US DOE grant No.
DE-FG02-97ER45639 (NEB) and NSF grant No. DMR-0225698 (KY).


\begin{references}
\bibitem{notation} It should be understood that all $SU(2)_k$ particles
considered here have topological charge 1/2.
\bibitem{noncrossing} This pair can always be brought together and
fused without any ``braiding" induced transitions because the
total topological charge of the particles between them is 0.
\bibitem{subtle} For $k\rightarrow \infty$, the alternate limit
$k\gg N\gg 1$ must be taken for which $S_A \simeq N  - \frac{3}{2}
\log_2 N$ and the entanglement entropy per bond is asymptotically
$1$, (consistent with $\log_2 d$ with $d=2$).  For ordinary
spin-1/2 particles $S_A$ is precisely $N$; the difference is due
to the lack of an $S_z$ degeneracy for $SU(2)_k$ particles.
\end{references}
\end{document}